\documentstyle[preprint,prb,aps]{revtex}
\begin{document}
\draft
\title{Quantum and classical phase transitions in
double-layer quantum Hall ferromagnets}
\author{Lian Zheng} 
\address{Department of Physics,
University of Maryland, College Park, Maryland 20742-4111}
\date{\today}
\maketitle
\begin{abstract}
We consider the problem of quantum and classical 
phase transitions in double-layer quantum Hall systems 
at $\nu=1/m$ ($m$ odd integers)
from a long-wavelength statistical mechanics viewpoint. 
We derive an explicit mapping of the
long-wavelength Lagrangian of the quantum Hall system
into that of a three-dimensional isotropic classical 
$XY$ model whose coupling constant depends 
on the quantum fluctuation in the original 
quantum Hall Hamiltonian.
Universal properties of
the quantum phase transition  
at the critical layer separation
are completely determined by this mapping.
The dependence of the 
Kosterlitz-Thouless transition temperature 
on layer separation, including quantum fluctuation effects,
is approximately obtained by simple finite-size
scaling analyses.
\end{abstract}
\narrowtext
%\section{}
Low-dimensional electron systems exhibit a richer variety of 
physical properties than their higher-dimensional  counterparts 
due to enhanced interaction effects.
For a two-dimensional electron gas in a perpendicular magnetic field,
the interaction effects are especially important 
because of Landau level quantization.
When electrons are entirely restricted to the lowest Landau level by a strong 
magnetic field, electron-electron interaction completely dominates 
the properties of the system as the electron kinetic energy
is quenched to an unimportant constant. 
One of the most interesting phenomena in these strongly correlated
electron systems is the quantum Hall effect (QHE), which has attracted 
a great deal of experimental and theoretical interest.\cite{qhe1}
In recent years, a lot of attention has been directed 
to quantum Hall systems in double-layer structures where 
electrons
are confined to two parallel planes separated by a distance 
comparable to the in-plane inter-electron distance.
With the introduction of this layer degree of freedom, 
many qualitatively new effects due entirely to interlayer electron 
correlations appear.\cite{mur1,kun1,wen1,nuh,pel1,me1,boe1,nu1}
These new features include 
QHE phases with various
spontaneously-broken symmetries, 
such as 
the interlayer coherent state
\cite{kun1} at $\nu=1/m$ 
($m$ odd integers)
and the canted antiferromagnetic state \cite{me1}
at $\nu=2$,
where interesting phase transitions both at zero
and finite temperatures may occur.  
Thus, multi-component quantum Hall systems 
provide a suitable platform for studying
various quantum phase transitions 
and their crossover behaviors. \cite{son1}
In this paper, we consider the 
quantum phase transition at $d=d_c$
and the Kosterlitz-Thouless transition at $d<d_c$
in double-layer systems at $\nu=1/m$, where 
$d$ is the layer separation.
Our consideration is based on
an explicit mapping of the long-wavelength 
Lagrangian of the quantum Hall system
into that of a three-dimensional (3D) isotropic 
classical $XY$ model whose coupling constant $g$ depends 
on the quantum fluctuation terms of the original Hamiltonian. 
The mapping shows unambiguously
that the
quantum phase transition 
at $d_c$ is in the same universality class as 
that of a 3D $XY$-model transition at 
its critical coupling constant $g_c$. 
The dependence of 
the Kosterlitz-Thouless transition temperature 
on layer separation
is approximately obtained by a straightforward
finite-size scaling analysis
around the quantum critical point.
In this way, 
both quantum and classical phase transitions in this problem
are described in terms of the known properties of 
a 3D classical $XY$ model.

To be specific, we restrict ourselves to $\nu=1$ ({\it i.e.} $m=1$), 
where various energy scales can be determined in 
the Hartree-Fock approximation.\cite{kun1} 
Our results, however,
apply qualitatively to the general case of $\nu=1/m$
with $m$ an odd integer. 
There has been a lot of work on the 
$\nu=1$ quantum Hall system.  
\cite{mur1,kun1,wen1,boe1,nu1}
At large layer separations, where the interlayer Coulomb interaction 
is negligible, the double-layer system is effectively 
a pair of decoupled half-filled
single layers 
which exhibit no QHE.   
At small layer 
separations, the interlayer Coulomb interaction is
almost as important as the intralayer interaction.
All electrons are in the symmetric state 
where interlayer and intralayer electron 
correlations are treated on an equal footing.
At $\nu=1$, the electrons form a filled band and exhibit the QHE.
The QHE phase at small $d$
and the non-QHE phase at large $d$ are separated by a continuous
transition 
at $d=d_c$.
For convenience,  we introduce a pseudospin variable $S$ to describe
the layer degree of freedom,\cite{kun1,nu1} where 
$S_z=\pm1/2$ represent electron occupation of the right or left layers,
respectively, and $S_x=\pm1/2$ represent electron occupation
of the symmetric or antisymmetric subbands, respectively. 
The transition between the QHE 
and non-QHE phases at $d=d_c$ may be viewed
as a magnetization transition:
The non-QHE phase at $d>d_c$ corresponds to a pseudospin disordered phase
and the QHE phase at $d<d_c$ corresponds to
a pseudospin magnetization
in $\hat x$-direction.
(Note that we assume that physical spins 
of the electrons are completely polarized
by the applied magnetic field and are not 
relevant variables at all.)
Even though, there has been a lot of work on the $\nu=1$ system,
most theoretical efforts,\cite{kun1,nu1} however, were 
directed towards the understanding
of the properties in the QHE phase. 
These studies usually 
ignore quantum fluctuations, and hence shed no light 
on the nature of the quantum phase transition at $d_c$.
In fact, many quantities obtained in this way,
such as the pseudospin stiffness, the magnetization,
and the susceptibility, do not show any sign of a phase transition
at all.
The present work is concerned solely with the phase transition, so it is 
essential that we   
include quantum fluctuations.
Our goal is accomplished by a mapping of the low energy physics
of the $\nu=1$ system into that of a 3D classical $XY$ model.
Although, the basic ideas involved here
have largely been known
in the literature,\cite{kun1,wen1} 
we think  
it is still very useful to explicitly carry out the derivations
and put these ideas into 
a concrete basis so that more sophisticated calculations may start
from here.

For the purpose of discussing the spontaneous pseudospin 
magnetization, we prohibit interlayer tunneling. (The tunneling
acts like a Zeeman term in the pseudospin space.) 
This is not an unreasonable restriction, as the interlayer tunneling
can be made very small in real semiconductor samples.
The Hamiltonian of the $\nu=1$ double-layer quantum Hall system is
\begin{eqnarray}
{\cal H}=&&{1\over2}\sum_{ij}
\sum_{\alpha_1\alpha_2}{1\over\Omega}\sum_{\bf q}
V_{ij}(q)e^{-q^2l_o^2/2}e^{iq_x(\alpha_1-\alpha_2)l_o^2}
\nonumber \\
&&\times C^\dagger_{i\alpha_1+q_y}C^\dagger_{j\alpha_2}
C_{j\alpha_2+q_y}C_{i\alpha_1},
\label{equ:h1}
\end{eqnarray}
where $\Omega$ is the area of the sample,
$l_o$ is the magnetic length,
and $C_{i\alpha}$ annihilates an electron in the lowest Landau
level
in layer $i$ ($i=1,2$) and
with the intra-Landau level index $\alpha$.
The interaction potentials are
$V_{ij}={2\pi e^2/\epsilon q}$ for $i=j$ and
$V_{ij}=V_{11} e^{-qd}$ for $i\ne j$.
At $d=0$, ${\cal H}$ has a $SU(2)$ symmetry in the pseudospin space.
When $d>0$, only a $U(1)$ symmetry is left because ${\cal H}={\cal H}(S_z)$.
Since $V_{11}>V_{12}$ for $d>0$, it is always energetically 
favorable to maintain
equal occupations of electrons in the two layers.
Therefore, ${\cal H}$ describes an easy-plane pseudospin magnetism
with possible spontaneous magnetization
only in the $\hat x$-$\hat y$ plane.
Since $[{\cal H}, S_{x,y}]\ne0$, 
the QHE phase, which has a pseudospin ordering in 
the $\hat{x}$-$\hat{y}$ plane,
is subjected to quantum fluctuations.

In the QHE phase, charge excitations \cite{kun1,son2} are gapped,
so the relevant degrees of freedom at low temperatures 
involve only neutral pseudospin excitations.
Then, the partition function can be expressed in terms of 
a coherent state path integral over pseudospin configurations
\begin{equation}
Z=\int_{|{\bf m}|=1} {\cal D}{\bf m}
\ e^{-{\cal S}_E({\cal \bf m})},
\label{equ:s1}
\end{equation}
where ${\bf m}$ is a unit vector representing the orientation 
of the pseudospin at position $({\bf r},\tau)$. The Euclidean action is
\begin{equation}
{\cal S}_E({\cal \bf m})=\int d^2r\int_0^{L_\tau} d\tau\left(
E({\bf m})-{i\nu\over4\pi l_o^2}{\bf A \cdot } 
\partial_\tau{\bf m}
\right),
\label{equ:s2}
\end{equation}
where $L_\tau=1/k_BT$, the inverse of temperature, 
is the system size in the (imaginary) time-direction. We shall first consider  
the case of zero temperature, so $L_\tau=\infty$.
The vector potential ${\bf A}$ accounts for
the Berry phase accumulated under time evolution of the pseudospins:
$\epsilon_{ijk}\partial A_k/\partial m_j=m_i$. 
In the long-wavelength limit,
the energy functional $E({\bf m})$ has been obtained \cite{kun1} from 
the microscopic Hamiltonian 
of Eqn. (\ref{equ:h1})
\begin{eqnarray}
E({\bf m})&&=
{\beta_m}m_z^2+{\rho^A\over2}(\nabla m_z)^2
+{\rho^E\over2}\left[(\nabla m_x)^2+(\nabla m_y)^2\right] \\
&&\approx{\beta_m}m_z^2+{\rho^E\over2}(\nabla\phi)^2, 
\label{equ:e1}
\end{eqnarray}
where $\beta_m$ and $\rho^{A(E)}$ are constants
which will be given below.
These terms  have clear physical meanings:
The gradient terms intend to maintain a 
pseudospin ordering: an exchange-induced pseudospin stiffness;
$\beta_mm_z^2$ intends
to suppress pseudospin polarization in $\hat{z}$-direction,
{\it i.e.,} it intends to maintain equal occupations of electrons
in the two layers to minimize Coulomb interaction energy.
In arriving at Eqn. (\ref{equ:e1}), we have parameterized ${\bf m}=
(\sqrt{1-m_z^2}\cos\phi,\sqrt{1-m_z^2}\sin\phi, m_z)$,
neglected $(\rho^A/2)(\nabla m_z)^2$ in comparison with  
$\beta_m(m_z)^2$ under the long-wavelength approximation,
and assumed $|m_z|\ll1$ because of the suppression of 
pseudospin polarization in $\hat{z}$-direction
at finite layer separations.
The $\beta_mm_z^2$ term in Eqn. (\ref{equ:e1}) 
carries the quantum fluctuation effects:
It is proportional to the expectation value
of $(\partial/\partial\phi)^2$, which clearly does not 
commute with $\phi$. This term is, for example, equivalent to 
the charging energy in Josephson junction arrays.

At $\nu=1$, we have \cite{kun1}
\begin{eqnarray}
\beta_m&&={\nu\over16\pi^2l_o^2}
\int_0^\infty q^2V_{11}(q)\left[d-(1-e^{-qd})/q\right]e^{-q^2l_o^2/2}dq,
\nonumber \\
\rho^E&&={\nu^2l_o^2\over32\pi^2}\int_0^\infty q^3
V_{11}(q)e^{-qd}e^{-q^2l_o^2/2}dq,
\label{equ:b1}
\end{eqnarray}
with $\rho^A=\rho^E(d=0)$.
$\rho^E$ decreases
as the layer separation increases:
It is the exchange-induced pseudospin stiffness associated with 
the interlayer phase 
coherence \cite{kun1}
in the $\nu=1$ system.
Contributions to $\beta_m$ come largely from the
static charging energy of the
double-layer electronic system.
$\beta_m$ increases monotonically
as a function of $d$ with 
$\beta_m(d=0)=0$, which represents the 
increased suppression of pseudospin polarization in $\hat{z}$-direction
as $d$ gets larger.
For $d$ not too small, we may think that the
low temperature physics is completely dominated by
the $\phi$-flucatuations and we may integrate out 
the $m_z$ degree of freedom: 
\begin{equation}
Z=\int_{|{\bf m}|=1}{\cal D}{\bf m}\ e^{-S^E({\bf m})} 
=\int{\cal D}\phi\
e^{-S_{\rm eff}(\phi)},
\label{equ:z2}
\end{equation}
with 
\begin{eqnarray}
e^{-S_{\rm eff}(\phi)}&&=\int{\cal D}m_z\ e^{-S^E({\bf m})} \nonumber \\
&&=\int{\cal D}m_z\
e^{-\int d^2r\int d\tau \left[
\beta_mm_z^2+{\rho^E\over2}(\nabla\phi)^2-{i\nu\over4\pi l_o^2}
{\bf A \cdot }\partial_\tau{\bf m}\right]}.
\label{equ:ze1}
\end{eqnarray}
The integration over $m_z$ is straightforward: 
$\phi$ and $m_z$ are coupled only through the Berry phase term
which, under a suitable gauge choice,
is ${\bf A \cdot }\partial_\tau{\bf m}=-m_z\partial_\tau\phi$.
Up to an irrelevant constant, we obtain
\begin{equation}
S_{\rm eff}(\phi)={1\over g}\int d^2r dx_0
\left[(\nabla\phi)^2+({\partial\over\partial x_0}\phi)^2\right],
\label{equ:se2}
\end{equation}
where the coupling constant is 
\begin{equation}
g={8\sqrt{2}\pi l_0^2\over\nu}\sqrt{\beta_m\over\rho^E},
\label{equ:g1}
\end{equation}
and the time-dimension has been rescaled by
\begin{equation}
\tau={\nu\over4\sqrt{2}\pi l_o^2\sqrt{\beta_m\rho^E}}x_0.
\label{equ:t1}
\end{equation} 

The action in Eqn. (\ref{equ:se2}),
$S_{\rm eff}$, is exactly that of 
a 3D classical $XY$ system,  which is known to have
a phase transition at  
$g=g_c=k_{3D}/\sqrt{2\pi l_o^2/\nu}$,
where $k_{3D}$ is the dimensionless critical coupling constant.
Since $g$ increases monotonically
as a function of $d$ with 
$g(d=0)=0$, the critical 
coupling constant $g_c$ thus corresponds to a critical 
layer separation $d_c$ 
given by $g(d_c)=g_c$, which is 
\begin{equation}
\left(\sqrt{l_o^2\beta_m\over\rho^E}\right)_{d=d_c}={1\over8 k_{3D}}
\sqrt{\nu\over\pi}.
\label{equ:dc1}
\end{equation}
We have succeeded in mapping the long-wavelength physics of the
quantum Hall system into that of an isotropic 3D classical $XY$ model.
The result shows that there is an easy-plane pseudospin
order-disorder transition at $g_c$, which is associated with the
QHE to non-QHE quantum phase transition at $d_c$. 
Universal properties of this
quantum phase transition  
are therefore the same as those of a 3D classical 
$XY$ model, which are well known.\cite{xy1}
The fact that $g\propto\sqrt{\beta_m}$ suggests that
the phase transition at $d_c$ is driven by quantum fluctuations.
Note, however, that the value of the critical layer separation $d_c$,
which is not a universal quantity, may not be accurately
given by this approach.
In practice, one may treat $g_c$, or $k_{3D}$,
as an adjustable parameter to make the value of $d_c$
given by Eqn. (\ref{equ:dc1}) match that found in experiments.

It is known \cite{kun1,wen1,nu1} that there is a linear mode 
in the QHE phase,
which is associated with the pseudospin-channel superfluidity.\cite{kun1} 
This neutral superfluid mode can be obtained easily in the present formalism:
A simple examination of Eqn. (\ref{equ:se2}) gives
\begin{equation}
\omega(q)={4\pi l_o^2\over\nu}\sqrt{2\beta_m\rho^E}\ q,
\label{equ:w1}
\end{equation}
which agrees completely with earlier results.\cite{kun1}
This mode is linear because the effective action $S_{\rm eff}$
in Eqn. (\ref{equ:se2}) is isotropic in the
three-dimensional $(\tau,{\bf r})$-space.

The zero-temperature easy-plane pseudospin order at $d<d_c$
persists, in the form of a quasi-long-range order,
up to a critical temperature.
This finite temperature transition is a  Kosterlitz-Thouless transition,
since the system is effectively
two-dimensional at $T>0$.
The critical temperature of the phase transition 
should depend on $d$ and vanish at $d=d_c$. 

For simplicity,  let us first consider the case where the
coupling constant $g$ in Eqn. (\ref{equ:se2})
is small so that the easy-plane pseudospin order can persist
up to relatively high temperatures where 
$L_0=(4\pi\sqrt{2}l_o^2/\nu)\sqrt{\beta_m\rho_E}L_\tau$, 
the system size
in $\hat{x}_0$-direction, is small.
We may neglect the time dependence of $\phi$,
{\it i.e.}, let $\phi(x_0,{\bf r})=\phi({\bf r})$,
and approximate Eqn. (\ref{equ:se2}) as
\begin{equation}
S_{\rm eff}(\phi)={L_0\over g}\int d^2r (\nabla\phi)^2,\ \ \ \ \
{\rm for}\ \ L_0\rightarrow0.
\label{equ:see1}
\end{equation}
It becomes a well-studied two-dimensional $XY$ model,
for which a Kosterlitz-Thouless transition occurs at \cite{kos1}
\begin{equation}
(L_0)_c\approx{g\over\pi}, \ \ \ \ \ {\rm for\ small}\ g,
\label{equ:tc1}
\end{equation}
where $(L_0)_c=L_0(T_c)$. The critical temperature determined in this way 
is $k_BT=(\pi/2)\rho^E$, a result obtained earlier.\cite{kun1}
This result neglected the time dependence in $\phi$
and hence excluded quantum fluctuations.  It is valid only for small 
$g$ and fails completely at $g\rightarrow g_c$, where
the pseudospin order is destroyed by
quantum fluctuations even at zero temperature.
Correction to the critical temperature from the quantum fluctuations
at $g\rightarrow g_c$ can be analyzed by finite-size scaling arguments.
For small $T$ and $(g-g_c)$, the free-energy density of
the system is proportional to a universal function $f(L_0/\xi)$,
where $\xi=\xi_o(1-g/g_c)^{-\overline\nu}$,
with $\overline{\nu}=2/3$, is the correlation length of
a 3D $XY$ model.\cite{xy1}
Finite temperature phase transition, which corresponds to a singularity
point $u^*$ 
in $f(u)$, occurs at $(L_0)_c/\xi=u^*$,
{\it i.e.,} at $(L_0)_c=u^*\xi_o(1-g/g_c)^{-\overline{\nu}}$. 
The coefficient $( u^*\xi_o)$ can be fixed by the limit of 
Eqn. (\ref{equ:tc1}). 
These considerations give the critical temperature of
the Kosterlitz-Thouless transition in the $\nu=1$
double-layer system as
\begin{equation}
k_BT_c={\pi\over2}\rho^E(1-g/g_c)^{2/3},
\label{equ:tc2}
\end{equation}
for $0<(1-g/g_c)<<1$.  
This simple relationship should be experimentally verifiable.

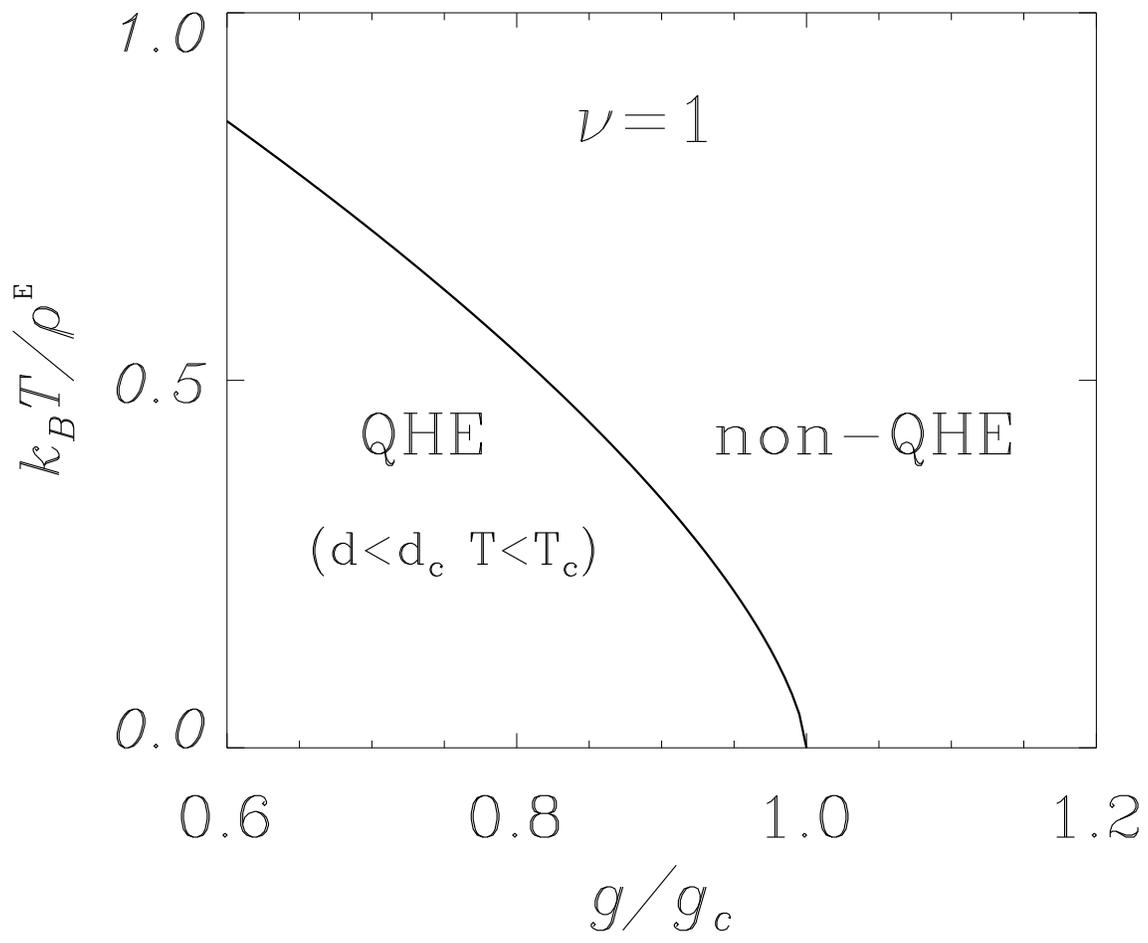
\begin{figure}
\caption{
Phase diagram of double-layer quantum Hall systems at
$\nu=1$, where $T$ is temperature and $g$ is the coupling constant.
The finite temperature phase boundary is given by
\protect{Eqn. (\ref{equ:tc2})}.
}
\label{fig1}
\end{figure}

Combining the descriptions
of both the zero temperature quantum phase transition
at $d=d_c$ and the finite temperature Kosterlitz-Thouless
transition at $d<d_c$,
a phase diagram for the
$\nu=1$ double-layer system
can be constructed 
and is shown in Fig. \ref{fig1}.  It is clear from the figure that
the QHE phase  
exists only for $d<d_c$ and $T<T_c$.
Experimental evidence for both
the quantum phase transition at $d=d_c$ and the
Kosterlitz-Thouless transition at $d<d_c$ 
has been observed.\cite{mur1,boe1}
The simple result of $T_c$ in Fig. \ref{fig1}, 
{\it i.e.} given by Eqn. (\ref{equ:tc2}),
also provides a useful check for any microscopic calculations of the
critical temperature.
We mention that the phase diagram given here
does not apply in the limit of $d\rightarrow0$ where out-of-plane
pseudospin fluctuations become important. 
This limit is, however, not experimentally accessible and is not
considered here.   

In summary, we have discussed both the zero temperature quantum phase 
transition at $d=d_c$ and the finite temperature
Kosterlitz-Thouless transition at $d<d_c$ in a $\nu=1$ double-layer
system by an explicit mapping of 
the long-wavelength Lagrangian
of the quantum Hall system into that of a 3D classical $XY$ model. 
The effects of quantum fluctuations are included naturally in our treatment.
This approach gives a simple and unified description of the 
quantum and classical phase transitions in terms 
of the known properties of the 3D $XY$ model. In particular,
it enables an approximate description of 
the effect of quantum fluctuations on the Kosterlitz-Thouless
transition temperature by finite-size scaling 
analyses. 
The results presented here are not quantitatively accurate, but they 
correctly capture
the essential physics of the phase transitions in double-layer quantum 
Hall systems at $\nu=1/m$ ($m$ odd integers).  

The author thanks Professor S. Das Sarma for
helpful discussions. This work is supported
by the US-ONR.

\end{document}